\documentclass{mn2e}
\usepackage{epsf}
\def\msun{{\rm M_{\odot}}}

\def\alf{\alpha_{0.1}}

\title[RS Ophiuchi: Thermonuclear Explosion or Disc Instability?] 
{RS Ophiuchi: Thermonuclear Explosion or Disc Instability?}

\author[King \& Pringle] {A.R. King$^{1}$ \& J.E. Pringle$^{1,2}$ \\
$^1$Theoretical Astrophysics Group, University of Leicester, Leicester
LE1 7RH\\
$^2$ Institute of Astronomy, Madingley Road, Cambridge, CB1 0HA}

\date{\today}

\volume{000}

\pagerange{\pageref{firstpage}--\pageref{lastpage}} \pubyear{2005}

\begin{document}

\label{firstpage}

\maketitle

\begin{abstract}
Sokoloski et al (2008) have recently reported evidence that the
recurrent nova RS Ophiuchi produced a pair of highly collimated radio
jets within days of its 2006 outburst. This suggests that an accretion
disc must be present during the outburst. However in the standard
picture of recurrent novae as thermonuclear events, any such disc must
be expelled from the white dwarf vicinity, as the nuclear energy yield
greatly exceeds its binding energy. We suggest instead that the
outbursts of RS Oph are thermal--viscous instabilities in a disc
irradiated by the central accreting white dwarf. The distinctive
feature of RS~Oph is the very large size of its accretion disc. Given
this, it fits naturally into a consistent picture of systems with
unstable accretion discs. This picture explains the presence and speed
of the jets, the brightness and duration of the outburst, and its rise
time and linear decay, as well as the faintness of the quiescence. By
contrast, the hitherto standard picture of recurrent thermonuclear
explosions has a number of severe difficulties. These include the
presence of jets, the faintness of quiescence, and the fact the the
accretion disc must be unstable unless it is far smaller than any
reasonable estimate.

\end{abstract}


\section{Introduction}

Sokoloski et al (2008) have presented evidence that within days of its
2006 outburst the recurrent nova RS Ophiuchi produced a pair of highly
collimated ('half--opening angles of a just few degrees') and high
velocity ($v \sim 5000$ km s$^{-1}$) jets. Taylor et al. (1989) made
similar deductions after the previous (1985) outburst. We argue here
that this provides compelling evidence about the cause of the outburst
in this object.

The source of the energy driving the outbursts in recurrent novae has
been under discussion for some time (e.g. Livio, Truran \& Webbink,
1986). There are two basic possibilities: {\it either} the outburst is
caused by a nuclear explosion on the surface of a white dwarf, akin to
a classical nova explosion, but smaller in scale, {\it or} the
outburst is powered by accretion energy released by a change in the
accretion rate on to the central white dwarf, akin to dwarf nova
outbursts but larger in scale, and similar in nature to the transient
X--ray binaries (which differ in having a central neutron star or
black hole as the accretor).

It is, however, difficult to see how a nuclear--powered explosion on
the surface of an accreting white dwarf could give rise to such
well--collimated jets. In general one would expect nuclear burning in
the hydrogen--rich surface shell to be quasi--spherically symmetric. The
nuclear energy yield from the hydrogen--rich matter considerably
exceeds its binding energy at the surface of a white dwarf. In a
nuclear--powered nova, the mass of explosively burning material is
comparable to that accreted since the last outburst. The mass in the
central regions of any disc must be much smaller than this, and is
even less gravitationally bound than the matter on the white dwarf
surface. Thus the mass in such a disc does not have enough inertia to
provide strong collimation of the flow. This point is underlined by
the lack of any strongly collimated flows to be seen in ejecta from
classical novae (Slavin et al., 1995; Gill \& O'Brien, 1998). One only
sees slight asymmetries in the shells of material expelled by the nova
explosion, presumably caused by the presence of a close binary
companion, and lumpiness in the shells (presumably caused by
instabilities in the ejection process).

In contrast, the jets seen in RS Oph are high velocity and strongly
collimated. As such they are strongly reminiscent of jets seen in
other astronomical objects such as radio galaxies, active galactic
nuclei (AGN), microquasars, some binary X--ray sources and young
stellar objects. These jets all have the following in common (Pringle,
1993; Livio, 1999, 2000; Price, Pringle \& King, 2003): the source of
the jet has accretion occurring through a disc; the jet velocities are
comparable to, or slightly higher than, the escape velocity from the
central accreting object; and about 10 per cent of the accreting
material is being ejected in the jet. There is also a suggestion that
the collimation of such jets require a large ratio of inner to outer
disc radii. It is immediately clear that the jets seen in RS Oph fall
precisely into this category.

We therefore argue here that we should abandon the view that the
recurrent outbursts of RS Oph are nuclear--powered. We propose instead
that the outbursts are driven by gravitational accretion energy, so
that the system is actually a dwarf nova with unusual properties which
all stem from the fact that its accretion disc is far larger in size
than in normal dwarf novae. This property leads to unusually bright,
long--lasting outbursts, in which irradiation of the disc by the
central source probably plays a role, and similarly long
quiescences. 
Many such outbursts are observed to have jets.

In the rest of this paper we first show that a dwarf nova origin offers a
reasonable explanation of the outbursts of RS Oph, and then examine the
evidence for a thermonuclear origin. Section 4 is a discussion.

\section{RS Oph as an irradiated dwarf nova}

The orbital period of RS Oph is 453.5 days (Brandi et al., 2009; Fekel
et al., 2000), implying a binary separation $a \sim 1.4$~AU. The
companion star is a red giant, and the measured period gives its
orbital velocity as $30M_1^{1/3}(1+M_2/M_1)^{-2/3}$~km\,s$^{-1}$ where
$M_1, M_2$ are the white dwarf and giant masses in $\msun$, and mass
transfer stability requires 
$M_2 \la M_1$. 
(Assuming that the emission
wings of the H$\alpha$ line trace the motion of the white dwarf,
Brandi et al (2009) give $M_1 = 0.59\sin^{-3}i, M_2 = 0.35\sin^{-3}i$, with
$i$ the inclination.) Mass transfer could potentially occur either
through capture of the red giant wind or Roche lobe overflow. However
since the stellar wind velocity $\sim 20$~km\,s$^{-1}$ (Bode \& Kahn,
1985, p.~206) is unlikely to exceed the giant's orbital velocity the
dynamics of the mass transfer must be similar to Roche lobe overflow
even if the giant's photosphere does not fill this lobe. The system
thus has an accretion disc with outer radius $\sim 70\%$ of the white
dwarf Roche lobe, hence size $\sim 0.5a \sim$ few~$\times
10^{12}$~cm. Unless the outer radius of the accretion disc is far
smaller than this estimate, i.e. $\sim 4 \times 10^{10}$~cm $\sim
2\times 10^{-3}a$, the gas there must be much cooler than the typical
temperature $T_H\sim 6500$~K for hydrogen ionization, implying that
the system must undergo disc instabilities.

A disc instability at radius $R \sim 10^{12}R_{12}$~cm has
characteristic viscous timescale $t_{\rm visc} \sim R^2/\nu$ with $\nu$ the
kinematic vicosity. We adopt the standard parametrization $\nu =
\alpha c_sH$, where $c_s$ is the local sound speed and $H$ the disc
scaleheight, with $\alpha$ a dimensionless parameter. In a thin,
fully--ionized disc, a wide range of observations give a typical range
$\alpha \sim 0.1 - 0.4$ (King et al., 2007). Accordingly for the hot,
outburst state of the disc in RS Oph we take the viscosity coefficient
as $\alpha_h = 0.1\alf$. Using eqs (5.49) of Frank et al., 2002, this
gives
\begin{equation}
t_{\rm visc} = 150\alf^{-0.8}R_{12}^{1.25}\dot M_{21}^{-0.3}~{\rm days}
\label{tvisc}
\end{equation}
where $R_{12}$ is the disc radius in units of $10^{12}$~cm, and $\dot
M_{21}$ the peak outburst accretion rate in units of
$10^{21}$~g\,s$^{-1}$. Immediately before the outburst begins, a
significant part of the disc must reach the maximum surface density
$\Sigma_{\rm max}(R)$ for a range of disc radii $R$, and one could in
principle estimate the unstable disc mass by integrating this
quantity. But $\Sigma_{\rm max}(R)$ depends on the {\it cool}--state
disc viscosity, which is essentially unknown, so we take instead
$\Sigma(R) = f\Sigma_{\rm min} = 8.25\times
10^2fR_{12}^{1.05}M_1^{-0.35}\alpha_h^{-0.8}$~g\,cm$^{-2}$. Here
$\Sigma_{\rm min}$ is the minimum surface density attained at the end
of the outburst, so that $f > 1$ (cf eqn (5.84) of Frank et al.,
2002). Integrating this over the disc surface we can estimate the
total disc mass $M_h$ before outburst. We approximate the local peak
accretion rate as $\dot M \simeq 2M_h/t_{\rm visc}$, as appropriate
for an exponential or linear viscous decay (see below) and using
(\ref{tvisc}) above we can now solve for $\dot M$ and $t_{\rm visc}$
as functions of $R_{12}$, i.e.
\begin{equation}
t_{\rm visc} = 127\alf^{-0.8}R_{12}^{0.48}M_1^{0.15}~{\rm days}
\label{tvisc2}
\end{equation}
and
\begin{equation}
\dot M_{\rm peak} = 2\times
10^{21}R_{12}^{2.57}M_1^{-0.5}f^{1.43}~{\rm g\, s^{-1}}
\label{mdot}
\end{equation}
with the consequence
\begin{equation}
\dot M_{\rm peak} = 5.5 \times
10^{20}t_{100}^{5.4}\alf^{4.3}M_1^{-1/3}f^{1.43}~{\rm g\, s^{-1}}
\end{equation}
where $t_{100}$ is $t_{\rm visc}$ in units of 100~days.

Before using these relations for RS Oph, we note that they give very
reasonable estimates for known cases of viscous decays if we take the
parameter $f = \Sigma_{\rm max}/\Sigma_{\rm min} \sim $ few, as
indeed suggested by numerical modelling of disc instabilities (see
Lasota, 2001) for a review). 

Thus for a superoutburst of an SU UMa system we take $M_1 = 0.7,
R_{12} = 2.5\times 10^{-2}$ and find a peak outburst accretion rate
$2\times 10^{17}f^{1.43}~{\rm g\, s^{-1}}$ (accretion luminosity $L
\simeq 10^{34}f^{1.43}~{\rm erg\, s^{-1}}$) and a decay timescale of
$\sim 14$~days.

For a neutron--star X--ray transient we take $R_{12} = 0.1, M_1 = 1.4$
and find a slightly super--Eddington luminosity $L \simeq 5\times
10^{38}f^{1.43}~{\rm erg\, s^{-1}}$ with $t_{\rm visc} \simeq
42$~days.

Both estimates, and others that one can make for e.g. black--hole
accretors in transients, agree well with observed systems, so we now
apply eqns (\ref{tvisc2}, \ref{mdot}) to RS Oph.

\subsection{Application to RS Oph}

For RS Oph we take $R_{12} \simeq 1$, $M_1 \simeq 1$ and find a peak
accretion rate 
\begin{equation}
\dot M_{\rm peak} \simeq 1.7\times
10^{21}f^{1.43}~{\rm g\, s^{-1}} = 2.5\times 10^{-5}\msun\, {\rm yr}^{-1},
\label{rsophacc}
\end{equation}
with an outburst timescale 
\begin{equation}
t_{\rm visc} \simeq 128\alf^{-0.8}~{\rm days}. 
\label{tvisc}
\end{equation}

The accretion rate (\ref{rsophacc}) is somewhat super--Eddington (for
gravitational energy release), formally giving a luminosity $GM\dot
M_{\rm peak}/R = 4.6\times 10^{38}$~erg\,s$^{-1}$ for accretion on to
a $1\msun$ white dwarf, so that $\dot M_{\rm peak} \simeq 3\dot M_{\rm
  Edd}$. In line with expectations from other super--Eddington
accretors such as ultraluminous X--ray sources we anticipate strong
outflow from the inner parts of the accretion disc, leading to an
accretion luminosity $L \simeq L_{\rm Edd}[1 + \ln(\dot M_{\rm
    peak}/\dot M_{\rm Edd})] \simeq 2L_{\rm Edd}$ (cf Shakura \&
Sunyaev, 1973; Begelman et al., 2006, Poutanen et al., 2007). At this
value $(\dot M_{\rm peak}/\dot M_{\rm Edd} \simeq 3$ the emission is
almost isotropic (at higher values of this ratio it is collimated
towards the disc axis). The system should thus have a luminosity of
$\sim 2L_{\rm Edd}$ unless viewed at very high inclination. This agrees
with observations of RS Oph, as does the outburst duration implied by
(\ref{tvisc}). 

The rise of RS Oph to maximum was very rapid ($\la 1$~day). This
follows from the dwarf nova picture given here also. To brighten from
an accretion luminosity of $10^{37}~{\rm erg\, s^{-1}}$ to
$10^{38}~{\rm erg\, s^{-1}}$ requires the disc instability heating
front to move from disc radius $R_d\sim 10^{10}$~cm to the white dwarf
radius, which is about 10 times smaller. Eqn (51) of Lasota (2001)
shows that the disc instability must be of `inside--out' type, as the
opposite would require a mass transfer rate greater than about
$4\times 10^{20}$~g\,s$^{-1}$, i.e. comparable with the outburst
accretion rate. For such outbursts the front moves with a velocity
$\la \alpha_h c_s$, so the rise time is
\begin{equation}
t_{\rm rise} \ga {R_d\over \alpha_h c_s}.
\end{equation}
Using the disc central temperature given by Frank et al., 2002 (eqn 5.49) to
estimate $c_s$ we find $t_{\rm rise} \ga 3$~hours.

Thus far the dwarf nova model gives a consistent representation of the
outburst of RS Oph. However we need one modification of the usual
picture. A normal dwarf nova outburst would end not after a viscous
time, but after a local thermal time, which is considerably
shorter. The crucial difference in RS Oph is the very large size of
the accretion disc. This means that irradiation by the central source
(temperature $\propto R^{-1/2}$) must dominate local viscous energy
release (temperature $\propto R^{-3/4}$) at large disc radii, where
most of the disc mass is. The outburst then resembles those of soft
X--ray transients: the disc is trapped in the hot outburst state until
central accretion drops because most of the heated mass has been
accreted (King \& Ritter 1998). For a disc heated by a central point
source of luminosity $L_c$ the irradiation temperature $T_{\rm
  irr}(R)$ is given by (cf King \& Ritter, 1998)
\begin{equation}
T_{\rm irr}^4 = {L_cg\over 4\pi R^2\sigma}\left({H\over R}\right)
\end{equation}
where $\sigma$ is the Stefan--Boltzmann constant, $g$ is a geometric
factor of order 0.2, and $H/R \sim 0.04$ is the disc aspect
ratio. This gives an irradiation temperature of order
\begin{equation}
T_{\rm irr} \simeq 6700\left({L_c\over L_{\rm
    Edd}R_{12}^2}\right)^{1/4}~{\rm K}
\end{equation}
that is, the central accreting white dwarf keeps the disc
self--consistently in the hot state ($T_{\rm irr} > T_H$) out to a
radius $R_h\sim 10^{12}$~cm at the start of the outburst. In this case
the outburst must evolve on the local hot--state viscous time, and
decay exponentially or linearly in time depending on whether
irradiation keeps the whole disc in the hot state or not (King \&
Ritter, 1998). As $R_h$ is smaller than the full disc outer radius,
the outburst decay should be (bolometrically) linear over a viscous
timescale. This is
consistent with observation
(Page et al., 2008, Fig.1) between days 50 to 100.

Between outbursts we expect the disc to refill, with little or no
accretion on to the white dwarf. This offers a natural explanation for
the faintness of the system between outbursts, in particular in
X--rays (Mukai 2009).  We conclude that the main features of the
outbursts of RS Oph are successfully explained in terms of an
instability in a disc irradiated by the central accreting white dwarf.

\section{RS Oph as a recurrent thermonuclear source}

We have suggested above that the outburst behaviour of RS Oph is
explicable in terms of instabilities in an irradiated disc. Until now
the usual interpretation of RS Oph and other recurrent novae is that
matter accumulates on the surface of the white dwarf until it reaches
the base pressure $P_{\rm crit} \sim 2 \times 10^{19}$~dyne~cm$^{-2}$
required to ignite as a runaway thermonuclear event. We note that the
presence of processed material in the outbursts is not per se
proof of nuclear burning in the outburst, since the companion star is
evolved. 
Thus any material transferred to the white dwarf and subsequently
expelled in an accretion--driven outburst came originally from the
convective envelope of the red giant, where it mixed with products of
the nuclear--burning shell around the degenerate core. Its composition
therefore does not differ greatly from that of matter partially burnt
and ejected from the white dwarf surface in a classical nova.

This picture has several difficulties, which we list below.

1. The presence of jets requires an accretion disc, and is very hard to
   reconcile with a thermonuclear model, as noted in the Introduction. The
   nuclear energy yield from burning hydrogen--rich matter considerably
   exceeds its gravitational binding energy at the surface of a white
   dwarf. Hence a thermonuclear explosion would inevitably blow away the inner
   parts of any accretion disc.

2. Even assuming a mass $M_1$ close to the Chandrasekhar limit, the
white dwarf has to accrete at a rate $\dot M = P_{\rm
  crit}R_1^4/GM_1t_{\rm qu} \ga 10^{-8}\msun~{\rm yr}^{-1}$ in order
to accumulate enough mass to ignite the outbursts (here $R_1$ is the
white dwarf radius, and $t_{\rm qu} \simeq 20$~yr is the duration of
quiesence). But the accretion rate $\dot M \ga 10^{-8}\msun~{\rm yr}^{-1}$
is hard to reconcile with the lack of X--rays observed from RS Oph in
quiescence. 
Mukai (2009) considers various possible ways out of this conclusion,
including combinations of very high intrinsic absorption columns and
totally optically thick boundary layers, and concludes that none are 
convincing. In particular, CVs with boundary layer emission always
in practice have a surface layer optically thin enough to produce hard
X--rays (Patterson \& Raymond, 1985), i.e. an optically thin region
where accretion energy is released.

3. The very wide binary orbit of RS Oph suggests that its disc is likely to
   be unstable, and thus reduce central accretion severely below the value 
   $\dot M \ga 10^{-8}\msun~{\rm yr}^{-1}$ required for the thermonuclear
   model. To avoid this the outer disc radius must be smaller than $\sim
   2\times 10^{-3}a$, where $a$ is the binary separation. This is ruled out if
   the giant fills its Roche lobe, and very unlikely even if mass transfer is
   via stellar wind capture, since the wind speed is rather less than the
   orbital velocity of the giant companion.

We note finally that the dwarf nova--type outbursts considered in
Section 2 imply a mean white dwarf accretion rate $\dot M_0 \la
7\times 10^{-8}\msun\, {\rm yr}^{-1}$ {\it if they recur at the
  current rate of one per $\sim 20$}~yr (note that mass loss during
the outburst makes this an upper limit). With a non--extreme white
dwarf mass one might then speculate on a thermonuclear nova occurring
after a few outbursts. However the ultimate source of mass is not the
disc, but the companion star. Unless this transfers mass at a rate
$\ga \dot M_0$, the disc will run out of mass and the outbursts will
recur more slowly. This is another way of saying that (as always) the
nova recurrence time is $ \sim \Delta M/(-\dot M_2)$, where $\Delta M$
is the mass that has to accumulate on the white dwarf surface. For
mass transfer on the nuclear timescale of a low--mass giant (as here),
the likely rate $-\dot M_2$ is less than $\dot M_0$ by factors $\ga 3$
(cf Webbink, Rappaport \& Savonije, 1983). This suggests a
`supercycle' of outbursts and a nova recurrence time $\ga
10^3$~yr. This is short compared with most CVs simply because $-\dot
M_2$ is much larger here.

\section{Discussion}

The distinctive feature of RS~Oph is the very large size of its
accretion disc. Given this, it fits naturally into a consistent
picture of systems with unstable accretion discs. This picture
explains the presence and speed of the jets, the brightness and duration
of the outburst, and its rise time and linear decay, as well as the
faintness of the quiescence. We would expect this general picture to
apply to other long--period systems, e.g. T CrB.

By contrast, the hitherto standard picture of recurrent thermonuclear
explosions has a number of severe difficulties. These include the
presence of jets, the faintness of quiescence, and the fact the
the accretion disc must be unstable unless it is far smaller than any
reasonable estimate.

A simple observational test of the thermonuclear picture would be
provided by a dynamical mass determination of the white dwarf mass. A
thermonuclear origin for the outbursts requires a mass extremely close
to the Chandrasekhar value $\sim 1.4\msun$.


\end{document}